\documentclass[12pt]{article}

\usepackage{amsfonts,amssymb,amsmath}
\usepackage[normalem]{ulem}

\DeclareMathAlphabet{\bfit}{OT1}{cmr}{bx}{it}

\DeclareFontFamily{U}{rsfs}{} \DeclareFontShape{U}{rsfs}{m}{n}{
  <5> rsfs5 <6> rsfs5 <7> rsfs7 <8> rsfs7
  <9> rsfs10 <10> rsfs10 <10.95> rsfs10
  <12-> rsfs10}{}
\DeclareMathAlphabet{\Cal}{U}{rsfs}{m}{n}

\def\mg{{\mathfrak g}}
\def\gl{{\Cal L}}

\newtheorem{thr}{Theorem}
\newtheorem{prop}{Proposition}
\newtheorem{cor}{Corollary}

\def\nn{ \nonumber }
\def\bq{ \begin{equation} }
\def\eq{ \end{equation} }
\def\ben{ \begin{eqnarray} }
\def\en{ \end{eqnarray} }

\def\on#1#2{\mathop{\vbox{\ialign{##\crcr\noalign{\kern2pt}
$\scriptstyle{#2}$\crcr\noalign{\kern2pt\nointerlineskip}
\kern-2pt$\hfil\displaystyle{#1}\hfil$\crcr}}}\limits}

\begin{document}

\title{Classical $R$-matrices for generalized $so(p,q)$ tops.}
\author{
A.V. Tsiganov\\
\it\small Department of Mathematical and Computational Physics,
 Institute of Physics,\\
\it\small St.Petersburg University,
 198 904, St.Petersburg, Russia.\\
\it\small e-mail: tsiganov@mph.phys.spbu.ru }

 \date{}
\maketitle

{\small An integrable deformation of the known integrable model of
two interacting $p$-dimensional and $q$-dimensional spherical tops is
considered. After reduction this system gives rise to the generalized
Lagrange and the Kowalevski tops. The corresponding Lax matrices and
classical $r$-matrices are calculated.}


\section{Introduction}
\setcounter{equation}{0}

The most common examples of classical $R$-matrices are associated
with decompositions of Lie algebra into a direct sum of two Lie
subalgebras \cite{rs03}. According to \cite{golsok} we can consider
various "perturbations" of the standard decompositions of the loop
algebras which may be associated with integrable deformations of the
known integrable systems \cite{sokts1,kst03}. In this note we
consider integrable deformations of some  tops associated with the
Lie algebra $so(p,q)$ and calculate the corresponding $R$-matrices.

Let ${\mg}$ be a self-dual Lie algebra, ${\mg}_{+},{\mg}_{-}\subset
{\mg}$ its two Lie subalgebras such that ${\mg}={\mg}_{+}\dot +%
{\mg}_{-}$ as a linear space. Let $P_+,\,P_-$ be the projection
operators onto $\mg_\pm$ parallel to the complimentary subalgebra;
the operator
\bq\label{r-mat} R=P_+-P_-
\eq
is a classical $R$-matrix. If $a$ is an intertwining operator
\bq\label{spl}
a[X,Y]=[aX,Y]=[X,aY]
\eq
for any $X,Y\in\mathcal G$, then  $R_a=R\circ a$ is also a classical
$R$-matrix \cite{rs03}. So, one decomposition of $\mg$ determines a
family of $R$-brackets those orbits are related
\bq\label{change-L}
L\mapsto a^{-1}L,
\eq
if $a$ is invertible. For instance, such change of variables was used
in \cite{rs03} to  connect Lax matrices for the Manakov top on
$so(4)$ and for the Clebsch system on $e(3)$.

The most interesting class of examples is provided by loop algebras
$\gl$
\bq\label{alg}
{\gl}=\mg[\lambda,\lambda^{-1}]=\left\{X(\lambda)= \sum_i x_{i}
\lambda^{i},\quad \quad x_{i}\in \mg\, \right\}.
\eq
 The standard decomposition $\gl=\gl_+\dot{+}\,\gl_-$ is defined by the natural $\mathbb Z$-grading in powers of auxiliare variable $\lambda$ (spectral parameter)
\bq\label{decomp-g}
 {\gl}_{+}= \bigoplus_{i\geq 0}
\,\mg_i\, \lambda^{i},\qquad {\gl}_{-}=\bigoplus_{i< 0} \,\mg_i\,
\lambda^{i}
\eq
The $R$-bracket associated with decomposition (\ref{decomp-g}) has a
large collection of finite-dimensional Poisson subspaces
\[\gl_{mn}=\oplus_{i=-m}^n\,\mg_i\,\lambda^i.\]
They are  $ad^*_R$-invariant subspaces
\bq\label{eq-Orb}
ad^*_RX\cdot L=(ad^*_\mg X_+\cdot L)_+-(ad^*_\mg X_-\cdot L)_-,
\qquad X_\pm=P_\pm X,\quad L\in\gl_{mn},
\eq
which are invariant with respect to the Lax equation
\bq\label{Lax-G}
L_t=[L,M],\qquad M=P_+(dH(L)).
\eq
Intertwining operators in $\gl$ are multiplication operators by
scalar Laurent polynomials. The crucial observation in
\cite{sokts1,kst03} is that to construct new Lax matrices by the rule
(\ref{change-L}) we can use the matrix Laurent polynomials $a$.

According to  \cite{golsok}, if $\mg$ is associative algebra, the
same space $\gl$ (\ref{alg}) is a linear sum of $\gl_+$
(\ref{decomp-g}) and subalgebra $\gl_-^y$
\bq\label{decomp}
\gl=\gl_+\dot{+}\,\gl_-^y,\qquad \gl_{-}^y=\left\{ \sum_{i=1} x_{i}
\lambda^{i} (1-\lambda^{-1} y) \quad \vert \quad x_{i}\in \mg,
\right\},
\eq
defined by some fixed element $y\in\mg$. This new decomposition of
$\gl$ defines  another classical $R$-matrix (\ref{r-mat})
\bq\label{Rg-y}
R_y=P_+^y-P_-^y.
\eq
Subalgebra $\gl_{-}^y$ could be regarded as some  "perturbation" of
the standard subalgebra $\gl_{-}$ by element $y$. In this approach
the finite-dimensional Poisson subspaces of the $R_y$-bracket
(\ref{Rg-y})
\[\bigoplus_{i=-m}^n\,\mg_i\,\lambda^i\,(1-\lambda^{-1} y)\qquad\mbox{\rm
or}\qquad (1-\lambda^{-1}
y)^{-1}\bigoplus_{i=-m}^n\,\mg_i\,\lambda^i,
\]
are deformations of the known orbits $\gl_{mn}$ of $R$-bracket, which
describe integrable deformations of the known integrable systems.

Our aim here is to show mapping of the  Lax matrices
\bq\label{in-orb}
L=\lambda{\bfit a}+{\bfit l} +\lambda^{-1} {\bfit s},\qquad M=\lambda
{\bfit b}
\eq
associated with Cartan type decomposition of $\mg=so(p,q)$ into the
new Lax matrices associated with $R_y$-bracket
\begin{equation}\label{LL}
L_y=(1-\lambda^{-1} y)^{-1} (\lambda  {\bfit a}+{\bfit
l}'+\lambda^{-1} {\bfit s}),\qquad M_y=M(1-\lambda^{-1}y),
\end{equation}
which describe new integrable deformations of $so(p,q)$ tops.


\section{Interacting spherical tops on $so(p,q)$ algebra}
\setcounter{equation}{0}

Let   $G=SO(p,q)$ is the group of pseudo-orthogonal matrices with
signature $(p,q)$, $p\geq q$. The Lie algebra $so(p,q)$ consists of
all the $(p+q)\times(p+q)$ matrices satisfying
\[X^T=-J\,X\,J,\]
where $J=\mbox{\rm diag}\,(1,\ldots,1;-1,\ldots,-1)$, tr$J=p-q$ and
$T$ means a matrix transposition.

In the natural $(p,q)$-block notation an element $X\in so(p,q)$ has
the form
\bq\label{sopq-m}
X=\left(%
\begin{array}{cc}
  \ell & s \\
  s^T & m \\
\end{array}%
\right),
\eq
where $\ell^T=-\ell$, $m^T=-m$ are $p\times p$ and $q\times q$
matrices, $s$ is an arbitrary $p\times q$ matrix .

The Cartan involution is given by $\sigma X=-X^T$ and $\mg=\mathfrak
f+\mathfrak p$ is the corresponding Cartan decomposition. The maximal
compact subalgebra $\mathfrak f=so(p)\oplus so(q)$ consists of
matrices $X$ with $s=0$
\[
\mathfrak f=so(p)\oplus so(q)=\left\{\left(%
\begin{array}{cc}
  \ell & 0 \\
  0 & m \\
\end{array}%
\right) \right\}
\]
The subspace $\mathfrak p$ consists of matrices $X$ with $\ell=0$,
$m=0$
\[
\mathfrak p=\left\{\left(%
\begin{array}{cc}
  0 & s \\
  s^T & 0 \\
\end{array}%
\right) \right\}.
\]
The pairing between $so(p,q)$ and $so(p,q)^*$ is given by invariant
inner product $(X,Y)=-\frac12\,\mbox{\rm trX}\,Y$ which is positive
definite on $\mathfrak f$.

We extend the involution $\sigma$ to the loop algebra
$\gl(\mg,\sigma)$ by setting $(\sigma X)(\lambda)=\sigma
(X(-\lambda))$. By definition, the twisted loop algebra
$\gl(so(p,q),\sigma)$ consists of matrices $X(\lambda)$ such that
\bq\label{X-pq}
X(\lambda)=-X^T(-\lambda).
\eq
The pairing between $\gl(\mg,\sigma)$ and $\gl(\mg,\sigma)$ is given
by $<X,Y>=\mbox{\rm Res }\lambda^{-1}(X,Y)$.


\subsection{Quadratic Hamiltonian}
Let $k\in SO(p)$, $r\in SO(q)$ and $\ell\in so(p)$, $m\in so(q)$ be
the configuration and momentum variables on the phase space
$T^*SO(p)\times T^*SO(q)$. All the non-zero Lie-Poisson brackets are
equal to
\begin{eqnarray}
\{\ell_{ij},\ell_{mn}\,\}&=&
\delta_{in}\,\ell_{jm}+\delta_{jm}\,\ell_{in}-\delta_{im}\,\ell_{jn}-\delta_{jn}\,\ell_{im}\,,\nn\\
\{\ell_{ij},k_{nm}\,\}&=&\delta_{jm}\,k_{ni}-\delta_{im}\,k_{nj}\,,
\nn\\
\label{Poi-Br}\\
\{m_{ij},m_{kn}\,\}&=&
\delta_{in}\,m_{jk}+\delta_{jk}\,m_{in}-\delta_{ik}\,m_{jn}-\delta_{jn}\,m_{ik}\nn\\
\{m_{ij},r_{nk}\,\}&=&\delta_{jk}\,r_{ni}-\delta_{ik}\,r_{nj} \,.\nn
\end{eqnarray}
The Poisson mapping from $T^*SO(p)\times T^*SO(q)$ into
$\gl(so(p,q),\sigma)$ is defined by the following theorem
\cite{rs03}.

\begin{thr} (Reyman, Semenov-Tian-Shansky)

The spectral invariants of the Lax matrix
\begin{eqnarray}\label{Lax-In}
L&=&\lambda\,{\bfit a}+{\bfit l}+\lambda^{-1}{\bfit s}\nn\\
\\
&=&\left(%
\begin{array}{cc}
  0 & k\,A\,r^T \\
  r\,A^T\,k^T & 0 \\
\end{array}%
\right)\lambda -
\left(%
\begin{array}{cc}
  k\ell k^T & 0 \\
  0 & r\,m\,r^T\\
\end{array}%
\right) +\lambda^{-1}\left(%
\begin{array}{cc}
  0 & F \\
  F^T & 0 \\
\end{array}%
\right),\nn
\end{eqnarray}
are in involution with respect to the canonical Poisson brackets
(\ref{Poi-Br}) on $T^*SO(p)\times T^*SO(q)$. Here $A$ and $F$ are
constant  $p\times q$ matrices.
\end{thr}
We refer the reader to \cite{rs03} for a complete proof of this
theorem which uses that $L(\lambda)$ (\ref{Lax-In}) is an orbit of
classical $R$-matrix (\ref{r-mat}) associated with a standard
decomposition of the twisted loop algebra $\gl(so(p,q),\sigma)$.

Here we present an elementary proof using a tensor form of the
$R$-bracket which is more familiar in the inverse scattering method
\cite{rs03}. According to \cite{bv90} functions $\phi_k(L)=\mbox{\rm
tr}\,L^k$, $k\geq 2$ are in the involution if and only if matrix
$L(\lambda)$ satisfies relation
\bq\label{rpoi}
\left\{\,{\on{L}{1}}(\lambda),{\on{L}{2}}(\mu)\,\right\}=
\left[r_{12}(\lambda,\mu),{\on{L}{1}}(\lambda)\right]
-\left[r_{21}(\lambda,\mu),{\on{L}{2}}(\mu)\,\right]\,.
\eq
Here  we used the standard notations
\[{\on{L}{1}}(\lambda)=
L(\lambda)\otimes 1,\qquad {\on{L}{2}}(\mu)=1\otimes L(\mu),
\]
and matrices $r_{12}$, $r_{21}$ are kernels of the operators $R$ and
$R^*$ such that
\[
r_{21}(\lambda,\mu)=\Pi\,r_{12}(\mu,\lambda)\,\Pi\,,
\]
where $\Pi$ is a permutation operator $\Pi X\otimes Y=Y\otimes X\Pi$.

One checks without difficulty that  $L(\lambda)$ (\ref{Lax-In})
satisfies (\ref{rpoi}) with the following kernel of the $R$-matrix
\bq\label{r12}
r_{12}(\lambda,\mu)=-\dfrac{\lambda^2}{\lambda^2-\mu^2}P_{\mathfrak
f} +\dfrac{\lambda\mu}{\lambda^2-\mu^2}P_{\mathfrak p},
\eq
where $P_{\mathfrak f}$ and $P_{\mathfrak p}$ are kernels of the
standard Casimir operators acting in the orthogonal subspaces
$\mathfrak f$ and $\mathfrak p$ respectively. They are equal to
\[P_{\mathfrak f}=\sum_{\alpha=1}^{\frac{p(p-1)+q(q-1)}{2}}
Z_\alpha\otimes Z_\alpha,\qquad P_{\mathfrak p}=\sum_{\beta=1}^{pq}
S_\beta\otimes S_\beta,
\]
where antisymmetric $Z_\alpha$ and symmetric $S_\beta$ matrices form
two orthonormal basises  in  $\mathfrak f$ and $\mathfrak p$. For
instance we can use
\[Z_1={ \left(%
\begin{array}{cccccc}
0 & 1 & 0 &\cdots & \vline&\\
  -1 & 0 & 0 & \cdots &\vline &0\\
   0& 0 & 0 &\cdots  &\vline&\\
   \vdots&\vdots&\vdots&\ddots&\vline&\\
   \hline
&  0 & & &\vline &  0
\end{array}%
\right)},\qquad
Z_2={ \left(%
\begin{array}{cccccc}
0 & 0 & 1 &\cdots & \vline&\\
  0 & 0 & 0 & \cdots &\vline &0\\
   -1& 0 & 0 &\cdots  &\vline&\\
   \vdots&\vdots&\vdots&\ddots&\vline&\\
   \hline
&  0 & & &\vline &  0
\end{array}%
\right)},\ldots\,,
\]
and
\[
S_1={ \left(%
\begin{array}{ccccccc}
0  & & &\vline& 1 & 0  &\cdots  \\

   & & &\vline &\vdots &\vdots&\ddots  \\
   \hline
1& 0 &\cdots &\vline & &  \\
\vdots& \vdots &\ddots &\vline & & 0
\end{array}%
\right)},\qquad S_2={
 \left(%
\begin{array}{ccccccc}
0  & & &\vline& 0 & 1  &\cdots  \\
   & & &\vline &\vdots &\vdots&\ddots  \\
   \hline
0& 0 &\cdots &\vline & &  \\
1& \vdots &\ddots &\vline & & 0
\end{array}%
\right)},\ldots.
\]

\begin{cor}
The equations of motion generated by  the spectral invariants of the
matrix  $L(\lambda)$ (\ref{Lax-In}) are Hamiltonian equations with
respect to canonical brackets (\ref{Poi-Br}), which give rise to Lax
equation (\ref{Lax-G}).
\end{cor}

Let $A$ be the truncated diagonal matrix $A=a_i\delta_{ij}$. The
spectral invariants $\phi_k(L)$ of  $L(\lambda)$ (\ref{Lax-In}) may
be combined in the following Hamiltonian
\ben\label{H_so_pq} H=\sum_{i<j\leqslant
q}\frac{a_ib_i-a_jb_j}{a_i^2-a_j^2} \,(\ell_{ij}^2+m_{ij}^2)
-2\sum_{i<j\leqslant q}\frac{a_ib_j-a_jb_i}{a_i^2-a_j^2}
\,\ell_{ij}m_{ij}+\nn\\
\nn\\
+\sum_{i=q+1}^p\sum_{j=1}^q \frac{b_j}{a_j}\,\ell_{ij}^2
+\gamma\sum_{i,j=q+1}^p \ell_{ij}^2
-2\sum_{i,j,s}b_ik_{ij}F_{js}r_{is},
\en
which formally describes an interaction of $p$-dimensional and
$q$-dimensional spherical tops. Here $a_i,b_i,~i=1\ldots q$ and
$\gamma$ are arbitrary parameters
 \cite{rs03}. In this case the second Lax matrix $M$ in
(\ref{Lax-G}) is equal to
\bq\label{M-gen}
M(\lambda)=P_+(dH(L))=-\left(%
\begin{array}{cc}
  0 & k\,B\,r^T \\
  r\,B^T\,k^T & 0 \\
\end{array}%
\right)\lambda,
\eq
where diagonal matrix $B=b_i\delta_{ij}$ includes  parameters $b_i$.


\subsection{Integrable deformations}
Let us introduce matrix
\bq\label{y1}
y=c\left(%
\begin{array}{cc}
  0 & F \\
  -F^T & 0 \\
\end{array}%
\right),\qquad c\in\mathbb C,
\eq
which is a generic solution of the following equations
\bq\label{cond}
 y\, X\, {\bfit s}=\sigma(y\, X\, {\bfit s})\,,\qquad
\forall X\in\mg,\qquad{\bfit s}=\left(%
\begin{array}{cc}
  0 & F \\
  F^T & 0 \\
\end{array}%
\right)\in \mathfrak p.
\eq
This solution is parametrized by an arbitrary numerical parameter
$c$.

The matrix $y$ (\ref{y1}), $y\in so(p+q)=\mathfrak f+i\mathfrak
p$, does not belong to the algebra $so(p,q)$ (\ref{sopq-m}). So,
to consider deformations $\gl_-^y$ (\ref{decomp}) of $\gl_-$ we
have to embed initial algebra $\mg$ and the element $y$ into some
algebra $\widetilde{\mg}$ and only then to discuss mapping of the
known orbit (\ref{in-orb}) of $R$-bracket (\ref{decomp-g}) into
the orbit (\ref{LL}) of $R_y$-bracket (\ref{Rg-y}) defined in
$\gl(\widetilde{\mg})^*$.

Therefore, for brevity, we shall consider kernels $r(\lambda,\mu)$
and $r^y(\lambda,\mu)$ instead of the corresponding operators $R$
(\ref{r-mat},\ref{decomp-g}) and $R_y$ (\ref{Rg-y},\ref{decomp}).

\begin{prop} \label{utv2}The
spectral invariants of the Lax matrix
\bq\label{Lax-y}
L_y=(1-\lambda^{-1} y)^{-1} (\lambda {\bfit a}+{\bfit
l}'+\lambda^{-1} {\bfit s}),
\eq
where
\bq\label{ncan-tr}
 {\bfit l}'={\bfit l}-\frac12\Bigl(y{\bfit a}+{\bfit a}y\Bigr),
\eq
are in involution with respect to the canonical Poisson brackets
(\ref{Poi-Br}) on $T^*SO(p)\times T^*SO(q)$.
\end{prop}
To proof this proposition we have to check equation (\ref{rpoi}) with
the kernel of the operator $R_y$
\bq\label{r-y}
r_{12}^{\,y}(\lambda,\mu)=\Bigl[1\otimes(1-\mu^{-1}y)^{-1}\Bigr]\,r_{12}(\lambda,\mu)\,
\Bigl[(1-\lambda^{-1}y)\otimes 1\Bigr].
\eq
by using properties of the element $y$ (\ref{cond}) and definition of
matrices $r_{ij}(\lambda,\mu)$ (\ref{r12}).

\begin{cor} The equations of motion generated by  the spectral invariants of
the matrix  $L_y$ (\ref{Lax-y}) are Hamiltonian equations with
respect to canonical brackets (\ref{Poi-Br}). They give rise to Lax
equation (\ref{Lax-G}), where
\bq\label{M-y}
{M}_y=P_+^y(dH(L))=M(1-\lambda^{-1}y).
\eq
\end{cor}
The Lax equation  (\ref{Lax-G}) for $L_y$ (\ref{Lax-y}) may be
rewritten as a Lax triad
\bq\label{Triada-1}
\dfrac{d}{dt}L_i=L_i\,M\,L_2- L_2\,M\,L_i,\qquad i=1,2
\eq
on the pair of matrices
\[L_1=L-\dfrac12(y{\bfit a}+{\bfit a}y)\qquad\mbox{\rm  and}\qquad
L_2=1-\lambda^{-1}y,
\]
entering in the definition of $L_y$ (\ref{Lax-y}).

For future reference we present two equivalent forms $H^{\, y}_{1,2}$
of the deformed Hamiltonian $H^{\, y}$ using additional canonical
transformations of the phase space
\bq\label{shift} {\bfit l}\to {\bfit
l}\pm\dfrac{c}2\,({\bfit s}{\bfit a}-{\bfit a}{\bfit s}).
\eq
Compositions of these canonical transformations with noncanonical map
(\ref{ncan-tr})
\[ {\bfit l}\to{\bfit l}_{1,2}={\bfit
l}-\dfrac12\Bigl((y\mp c{\bfit s}){\bfit a}+{\bfit a}(y\pm c{\bfit
s})\Bigr)
\]
act either in the subalgebra $so(p)$
\bq\label{l-shift} \ell\to
\ell_1=\ell-c\,\Bigl(k^T\,F\,r\,A^T-A\,r^TF^Tk\Bigr)\,,
\eq
either in the subalgebra $so(q)$
\bq\label{m-shift}
m\to m_1=m-c\,\Bigl(r\,A^Tk^TF-F^T\,k\,A\,r^T\Bigr)\,.
\eq
The corresponding perturbations of the initial Hamiltonian $H$
(\ref{H_so_pq}) depend either on $\ell$ variables
\bq\label{def-Hg} {H}_1=H-c\cdot{\rm
tr}\,(k\,\ell\,B\,r^T\,F^T)+\frac{c^2}{2}\,{\rm
tr}\,(B^TA\,r\,F^TF\,r^T)
\eq
either on $m$ variables
\bq\label{def-Hg2}
{H}_2=H+c\cdot{\rm tr}\,(r\,m\,B^T\,k^T\,F)+\frac{c^2}{2}\,{\rm
tr}\,(B\,A^T\,k^T F\,F^T k)\,.
\eq
The corresponding Lax matrices are equal to
\bq\label{Lax-y1}
L_y^{(1)}=(1-{\lambda}^{-1}y)^{-1}\left[L(\lambda)+c\left(%
\begin{array}{cc}
  FrA^Tk^T-kAr^TF^T & 0 \\
  0 & 0 \\
\end{array}%
\right)\right]
\eq
or
\bq\label{Lax-y2}
L_y^{(2)}=(1-{\lambda}^{-1}y)^{-1}\left[L(\lambda)-c\left(%
\begin{array}{cc}
  0 & 0 \\
  0 & FrA^Tk^T-kAr^TF^T \\
\end{array}%
\right)\right].
\eq

We have discussed so far Lax matrices for the tops in the stationary
reference frame, i.e. used Euler-Lagrange description of motion.
According to \cite{rs03}, we can go over to  the Lax matrices in the
frame moving with the body which amounts to the gauge transformation
\ben
\widetilde{L}(\lambda)&=&{\rm g}^TL{\rm g}= \lambda \tilde{\bfit
a}+\tilde{\bfit l}+\lambda^{-1} \tilde{\bfit s}= \nn\\ \nn\\
&=&\left(%
\begin{array}{cc}
  0 & A \\
  A^T & 0 \\
\end{array}%
\right)\lambda -
\left(%
\begin{array}{cc}
  \ell & 0 \\
  0 & m\\
\end{array}%
\right) +\lambda^{-1}\left(%
\begin{array}{cc}
  0 & k^T\, F\, r \\
 r^T\, F^T\,k & 0 \\
\end{array}%
\right),\nn\\
\label{Lax-st}\\
\widetilde{M}(\lambda)&=&{\rm g}^TM(\lambda){\rm g}+w=-\left(%
\begin{array}{cc}
  0 & B \\
  B^T & 0 \\
\end{array}%
\right)\lambda+{\rm g}^T\dfrac{d}{dt}{\rm g}.\nn
\en
Here  $ {\rm g}=\left(%
\begin{array}{cc}
  k & 0 \\
  0 & r \\
\end{array}%
\right)$ and $w={\rm g}^T\dot{{\rm g}}$ is the block-matrix of
angular velocities, related to $\ell$ and $m$ by a linear
transformations.

\begin{prop}
The Lax matrix $\widetilde{L}(\lambda)$ (\ref{Lax-st}) associated
with the Euler-Poison description of the motion satisfies equation
(\ref{rpoi}) with the following matrices
$\widetilde{r}_{ij}(\lambda,\mu)$
\bq
\widetilde{r}_{12}(\lambda,\mu)=-r_{12}(\lambda^{-1},\mu^{-1}),\qquad
\widetilde{r}_{21}(\lambda,\mu)=-r_{21}(\lambda^{-1},\mu^{-1}),
\eq
which are obtained from $r_{12}(\lambda,\mu)$ (\ref{r12}) by change
of the spectral parameters.
\end{prop}

Using the similar gauge transformation to the matrices $L_y(\lambda)$
(\ref{Lax-y}) and $M_y(\lambda)$ (\ref{M-y}) one gets
\ben
\widetilde{L}_y(\lambda)&=&{\rm g}^TL_{y}{\rm
g}=(1-\lambda^{-1}\tilde{y})^{-1} \left(
\widetilde{L}(\lambda)-\dfrac12(\tilde{y}\,\tilde{\bfit
a}+\tilde{\bfit a}\,\tilde{y})
\right),\nn\\
\label{Lax-sty}\\
\widetilde{M}_y(\lambda)&=&{\rm g}^TM_y(\lambda){\rm g}+{\rm
g}^T\dfrac{d}{dt}{\rm
g}=\widetilde{M}(1-\lambda^{-1}\widetilde{y})\,.\nn
\en
Here matrices $w_y={\rm g}^T\dot{{\rm g}}$ are calculated with
respect to cubic Hamiltonian $H^{y}$ and, therefore, it depends on
all the dynamical variables. As above (\ref{Triada-1}), a pair of
matrices
\[L_1=\widetilde{L}+\dfrac12(\tilde{y}\widetilde{\bfit
a}+\widetilde{\bfit a}\tilde{y})\qquad\mbox{\rm  and}\qquad
L_2=1-\lambda^{-1}\tilde{y},
\]
entering in the definition $\widetilde{L}_y$ (\ref{Lax-sty}), satisfy
to the Lax triad
\bq\label{Triada-2}
\dfrac{d}{dt}L_i(\lambda)=L_i(\lambda)\,\widetilde{M}(\lambda)+
\widetilde{M}\,^T(-\lambda)\,L_i(\lambda),\qquad i=1,2.
\eq
These equations (\ref{Triada-2}) have a more complicated structure
with respect to Lax triad in the rest frame (\ref{Triada-1})

In the body frame  the matrix
 \[\tilde{y}={\rm g}^T\,y\,{\rm
g}=c\left(%
\begin{array}{cc}
  0 & k^T\, F\, r \\
 -r^T\, F^T\,k & 0 \\
\end{array}%
\right)
\]
depends on the dynamical variables and, therefore, the kernel
(\ref{r-y}) of the corresponding operator $R_y$  depends on these
variables too.

However in the moving frame we can construct another Lax matrix such
that the kernel of the corresponding $R$-matrix does not depend on
dynamical variables. Namely, using matrix
\bq\label{z1}
\tilde{z}=c\left(%
\begin{array}{cc}
  0 & A \\
  -A^T & 0 \\
\end{array}%
\right)\eq such that\bq\tilde{z}\, X\, {\bfit a}=\sigma(\tilde{z}\,
X\, {\bfit a})\,,\qquad \forall X\in\mg,\qquad{\bfit
a}=\left(\begin{array}{cc}
  0 & A \\
  A^T & 0 \\
\end{array}%
\right)\in \mathfrak p,
\eq
we can consider perturbation  of the Lax matrix
$\widetilde{L}(\lambda)$ (\ref{Lax-st}) by $\tilde{z}$
\bq\label{Lax-stz}
\widetilde{L}_z(\lambda)=(1-\lambda \tilde{z})^{-1} \left(
\widetilde{L}(\lambda)-\dfrac12(\tilde{z}\,\tilde{\bfit
s}+\tilde{\bfit s}\,\tilde{z}) \right).
\eq
\begin{prop} The Lax matrix $\widetilde{L}_z(\lambda)$
(\ref{Lax-stz}) satisfies equation (\ref{rpoi}) with the following
numerical  $r$-matrix
\bq\label{r-z}
\widetilde{r}_{12}^{\,z}(\lambda,\mu)=\Bigl[1\otimes(1-\lambda\tilde{z})^{-1}\Bigr]\,\widetilde{r}_{12}(\lambda,\mu)\,
\Bigl[(1-\mu\tilde{z})\otimes 1\Bigr].
\eq
\end{prop}
The proof is straightforward.

Applying inverse gauge transformation to $\widetilde{L}_z(\lambda)$
one gets another Lax matrix $L_z(\lambda)$ in the stationary frame
\bq\label{Lax-z}
 L_z(\lambda)={\rm g}L_z{\rm g}^T=(1-\lambda z)^{-1} \left(
L(\lambda)-\dfrac12(z\,{\bfit s}+{\bfit s}\,z) \right),\qquad z={\rm
g}\,\tilde{z}\,{\rm g}^T.
\eq
The corresponding $r$-matrix is dynamical.

So, for initial Hamiltonian  $H$ (\ref{H_so_pq}) we know two Lax
matrices $L(\lambda)$ (\ref{Lax-In}) and $\widetilde{L}(\lambda)$
(\ref{Lax-st}) associated with the physically different coordinate
systems. For these Lax matrix we constructed by two perturbations
$L_y$ (\ref{Lax-y}), $L_z$
 (\ref{Lax-z}) in the rest frame and
$\widetilde{L}_y$ (\ref{Lax-sty}), $\widetilde{L}_z$
 (\ref{Lax-stz}) in the body frame.


\section{Further examples}
\setcounter{equation}{0}

According to \cite{rs03}, we shall try to exclude the nonphysical
degrees of freedom using symmetry of the Hamiltonian $H$
(\ref{H_so_pq}) and its perturbations $H_{1,2}$
(\ref{def-Hg},\ref{def-Hg2}). Assume that $A=E$ is the truncated
identity matrix $E_{ij}=\delta_{ij}$.

Below all the Hamiltonians will be expressed through kinetic momentum
$\ell_{ij}\in so(p)$  and the entries of the Poisson vectors
$x_i=k^T\,f_i$, where $f_i$ are the column vectors of the matrix $F$.
These variables are canonical coordinates on the algebra
$e(p,q)^*=so(p)\ltimes\mathbb (R^p\otimes\mathbb R^q)$ \cite{rs03}.


\subsection{Algebra $so(p,1)$, Lagrange top and spherical pendulum.}

If $q=1$ the phase space is $T^*SO(p)$. We can put
\bq \label{rest-Lag}
m=0,\qquad r=1,
\eq
without loss of generality. In this particular case the deformations
(\ref{def-Hg}) and (\ref{def-Hg2}) of the initial Hamiltonian are
quadratic polynomials instead of cubic ones.

Following to  \cite{rs03}, we can rewrite Hamiltonian (\ref{H_so_pq})
in the form
\bq \label{H-lagr}
H=\frac12\sum_{i,j=1}^p{\ell}^2_{ij}+
\frac{\gamma}2\sum_{i,j=2}^p{\ell}^2_{ij} - (e_1,x),\qquad x=k^T\,f.
\eq
Here $e_1$ is a first vector of the standard  basis  in $\mathbb
R^p$.

The Hamiltonian (\ref{H-lagr}) describes rotation of a rigid body
around a fixed point in a homogeneous gravity field. The vector $x$
is the vector along the gravity field, with respect to the body frame
and $e_1$ is the vector pointing from the fixed point to the center
of mass of the body.

It is Lagrange case because the body is rotationally symmetric and
the fixed point lies on the symmetry axis $e_1$. The perturbations
(\ref{def-Hg}) and (\ref{def-Hg2}) of the Hamiltonian (\ref{H-lagr})
\[{H}_1=H+\frac{c^2}2\,(e_1,\,k^T\,f)^2=H+\frac{c^2}2\,(e_1,x)^2
\]
\[\label{H_lag2}
 {H}_2=H+c\,(\ell\,e_1,\,k^T\,f)=H+c\,(\ell\,e_1,x).
\]
are quadratic polynomials and, therefore, the corresponding equations
of motion have the form of the Kirchhoff equations.

Let $x=k^Tf$ and $\pi$ are canonical coordinates in $\mathbb R^{2p}$,
$\{\pi_i,x_j\}=\delta_{ij}$. Substituting
$\ell_{i,j}=x_i\pi_j-\pi_ix_j$ in $L(\lambda)$ (\ref{Lax-In}) one
gets a Lax matrix for the spherical pendulum in appropriate physical
coordinates \cite{rs03}. The corresponding Hamiltonian
\[
H=\frac12\,\sum \pi_i^2-\sum A_ix_i.
\]
describes a motion on the sphere $S^{p-1}$, $(x,x)=1$, $(x,\pi)=0$,
in a homogeneous gravity field. Its deformations (\ref{def-Hg2}) and
(\ref{def-Hg}) look like
\[{H}_1=H+c\sum A_i\pi_i\,,\qquad
{H}_2=H+\frac{c^2}2\left(\sum A_ix_i\right)^2\,.
\]


\subsection{The Kowalevski top.}

According \cite{rs03,brs89}, the generalized $p$-dimensional
Kowalevski top is the reduced system (\ref{H_so_pq}) with respect to
the action of the subgroup $SO(q)$. The reduction amounts to imposing
the constraints
\bq\label{red-HK}
m+P\,\ell\,P=0,\qquad r=1.
\eq
Here $P$ means the orthogonal projection from $\mathbb R^p$ onto
subspace of $\mathbb R^q$ spanned by first $q$ vectors of the
standard basis. The reduced phase space is $T^*SO(p)$ with its
canonical Poisson structure.

Inserting the constraints (\ref{red-HK}) into (\ref{Lax-In}) one gets
the Lax matrix for the generalized Kowalevski top
\bq\label{Lax-Kow}
L^{Kow}(\lambda)=\left(%
\begin{array}{cc}
  0 & k\,E \\
  E^T\,k^T & 0 \\
\end{array}%
\right)\lambda -
\left(%
\begin{array}{cc}
  k\ell k^T & 0 \\
  0 & -E^T\,\ell\,E\\
\end{array}%
\right) +\lambda^{-1}\left(%
\begin{array}{cc}
  0 & F \\
  F^T & 0 \\
\end{array}%
\right).
\eq

\begin{prop} The spectral invariants of the Lax matrices $L^{Kow}(\lambda)$
(\ref{Lax-Kow}) and $L_y(\lambda)$ (\ref{Lax-y}) by (\ref{red-HK})
are in the involution  with respect to Lie-Poisson brackets on
$T^*SO(p)$.
\end{prop}
We present a direct proof of the involutivity of invariants without
recourse to Hamiltonian reduction.

In the rest frame it can easily be shown that  the reduced Lax
matrices  satisfy equation (\ref{rpoi}) with the same kernels
$r_{12}(\lambda,\mu)$ (\ref{r12}) and $r^{\,y}_{12}(\lambda,\mu)$
(\ref{r-y}). These kernels do not change by the Poisson reduction
(\ref{red-HK}) and the corresponding operators $R$ (\ref{r-mat}) and
$R_y$ (\ref{Rg-y}) remain differences of the same projectors.

It should be pointed out that constraints (\ref{red-HK}) agree with
the Euler-Lagrange description of the top, but disagree with its
Euler-Poisson description \cite{rs03}. In the body frame the Lax
matrix is given by
\bq\label{Lax-Kowb}
\widetilde{L}^{Kow}(\lambda)=\left(%
\begin{array}{cc}
  0 & E \\
  E^T& 0 \\
\end{array}%
\right)\lambda -
\left(%
\begin{array}{cc}
  \ell  & 0 \\
  0 & -E^T\,\ell\,E\\
\end{array}%
\right) +\lambda^{-1}\left(%
\begin{array}{cc}
  0 & k^TF \\
  F^Tk & 0 \\
\end{array}%
\right).
\eq
\begin{prop} The spectral invariants of the Lax matrices
$\widetilde{L}^{Kow}(\lambda)$ (\ref{Lax-Kowb}) and
$\widetilde{L}_z(\lambda)$ (\ref{Lax-stz}) by (\ref{red-HK}) are in
the involution with respect to Lie-Poisson brackets on $T^*SO(p)$.
\end{prop}
 In the body frame reduction (\ref{red-HK}) does not Poisson mapping
 which changes $R$-brackets. Thus, after reduction the Lax matrix
  $\tilde{L}(\lambda)$ (\ref{Lax-st})
 satisfies (\ref{rpoi}) with the reduced kernel
\bq\label{r-kow}
\widetilde{r}_{12}^{\,Kow}(\lambda,\mu)=\widetilde{r}_{12}(\lambda,\mu)-\sum_{\alpha=1}^{\frac{p(p-1)+q(q-1)}{2}}
Z_\alpha\otimes \widehat{P}Z_\alpha \widehat{P}\,.
\eq
Here
\[\widehat{P}=\left(%
\begin{array}{cc}
  0 & E \\
  E^T & 0 \\
\end{array}%
\right)\,,\qquad
\widehat{P}Z_\alpha \widehat{P}=\left\{%
\begin{array}{ll}
    Z_{\alpha\,+\frac{p(p-1)}2}, & 0<\alpha\leq \frac{q(q-1)}2; \\
    \\
    0, & \frac{q(q-1)}2<\alpha\leq \frac{p(p-1)}2; \\
    \\
    Z_\alpha, & \frac{p(p-1)}2<\alpha.
\end{array}%
\right.
\]
Algebro-geometric description of the corresponding "non-standard" $R$
operator may found in  \cite{mar98}.

After reduction (\ref{red-HK}) the perturbed  Lax matrix
$\widetilde{L}_z(\lambda)$ (\ref{Lax-stz}) satisfies (\ref{rpoi})
with the reduced $r$-matrix
\[\widetilde{r}_{12}^{\,zKow}(\lambda,\mu)=\Bigl[1\otimes(1-\lambda\tilde{z})^{-1}\Bigr]\,
\widetilde{r}_{12}^{Kow}(\lambda,\mu)\, \Bigl[(1-\mu\tilde{z})\otimes
1\Bigr].
\]

Substituting constraints (\ref{red-HK}) into (\ref{H_so_pq}) gives
the reduced Hamiltonian on $T^*SO(p)$
\bq\label{Kow-Ham}
H=\frac14\,\left(\sum_{i,j=1}^p \ell_{ij}^2+\sum_{i,j=1}^q
\ell_{ij}^2\right)- \sum_{i=1}^q (e_i,\,k^T\,f_i),
\eq
where $e_1,\dots ,e_p$ is the moving orthonormal frame associated
with the top, and  $f_1,\dots,f_q$ are the column vectors of $F$.
This Hamiltonian describes a $p$-dimensional top with a very special
$SO(q)\times SO(p-q)$-symmetric inertial tensor, rotating under the
influence of $q$ constant forces \cite{rs03}.

Inserting constraints (\ref{red-HK}) into ${H}_1$ (\ref{def-Hg}) and
${H}_2$ (\ref{def-Hg2}) one gets quadratic polynomials
\bq\label{def-Kowpq2}
{H}_1=H-c\,\sum_{i=1}^q
(P\,\ell\,P\,e_i,\,k^Tf_i)+\frac{c^2}2\sum_{i=1}^q (k^Tf_i,k^Tf_i)\,,
\eq
and
\bq\label{def-Kowpq}
{H}_2=H-c\,\sum_{i=1}^q (\ell\,e_i,\,k^Tf_i)\,,
\eq
It means that the corresponding equations of motion have the form of
the Kirchhoff equations of motion of a rigid body in the ideal fluid.

Thus, the characteristic doubling of the terms $\ell_{ij}^2$, for
$i,j\leqslant q$ in (\ref{Kow-Ham}), and existence of the nontrivial
quadratic perturbations ${H}_1^{\,y}$ (\ref{def-Kowpq2}) and
${H}_2^{\,y}$ (\ref{def-Kowpq}) are results of reduction with respect
to $SO(q)$.

The case  $p=3$, $q=2$ and $f_2=0$ gives usual Kowalevski top on the
algebra $e(3)$ with a Poisson vector $x=k^T\,f_1$. The deformations
(\ref{def-Hg}-\ref{def-Hg2}) of the standard Hamiltonian became
\bq \label{sHam}
{H}_1=\frac12\Bigl(\,\ell_{23}^2+\ell_{13}^2+2\ell_{12}^2\,\Bigr)-x_1+c\,\ell_{12}x_2
+\frac{c^2}2\,(x_1^2+x_2^2\,) \,.
\eq
and
\bq \label{sHam2}
{H}_2=\frac12\Bigl(\,\ell_{23}^2+\ell_{13}^2+2\ell_{12}^2\,\Bigr)
-x_1+c\,(\ell_{12}x_2+\ell_{13}x_3\,).
\eq
Recall, ${H}_1$ and ${H}_2$ are two different forms of the one
Hamiltonian $H^y$, which are connected by canonical transformation
(\ref{shift}).


\subsection{Integrable systems on $so(4)$.}
If $p=q=3$ and $F_{ij}=0$ the Hamiltonian $H$ (\ref{H_so_pq})
describes integrable system on the phase space $so(4)=so(3)\oplus
so(3)$. Let
\[
u_k=-\varepsilon_{ijk}\ell_{jk},\qquad v_k=-\varepsilon_{ijk}m_{jk},
\]
where $\varepsilon_{ijk}$ is the totally skew-symmetric tensor. The
Hamiltonian (\ref{H_so_pq}) becomes
\[ H=\sum_{k=1}^3\alpha_k(u_k^2+v_k^2)
-2\sum_{k=1}^3\beta_k\,u_kv_k.
\]
The coefficients
\[
\alpha_k=\varepsilon_{ijk}^2\frac{a_ib_i-a_jb_j}{a_i^2-a_j^2},\qquad
\beta_k= \varepsilon_{ijk}^2\frac{a_ib_j-a_jb_i}{a_i^2-a_j^2}\,
\]
satisfy the relation
\[
(\alpha_1-\alpha_2)\beta_3^2
     +(\alpha_2-\alpha_3)\beta_1^2
    +(\alpha_3-\alpha_1)\beta_2^2
+(\alpha_1-\alpha_2)(\alpha_2-\alpha_3)(\alpha_3-\alpha_1)=0.
\]
Thus, according to \cite{ves}, this integrable system belongs to the
so-called Steklov-Manakov family of integrable systems, characterized
by the property that there exists an additional quadratic integral.
In this case $F=0$ and proposed deformations
(\ref{def-Hg}-\ref{def-Hg2}) are trivial.

The more complicated passage to the algebra $so(4)$ was proposed in
\cite{kst03}. Let us apply composition of the gauge transformations
to the Lax matrix $L_y^{(2)}$ (\ref{Lax-y2})
\[
\hat{L}_y=(1-\lambda^{-1}g)\,\Bigl(\,{\rm
g}^T\,L_y^{(2)}(\lambda)\,{\rm g}\,\Bigr)\,(1-\lambda^{-1}g)^{-1},
\qquad g=c\left(%
\begin{array}{cc}
  0 & k^TF \\
  0 & 0 \\
\end{array}%
\right)\,.
\]
Substituting constraints (\ref{red-HK}) one gets
\ben
\hat{L}_y=&&\left[1+\dfrac{c^2}{\lambda^2}\left(%
\begin{array}{cc}
  0 & 0 \\
  0 & F^T\,F \\
\end{array}%
\right)\right]^{-1}\times\nn\\
 \label{Lax-hat}\\
 \nn\\
 &&\left[
\left(%
\begin{array}{cc}
  0 & E \\
  E^T& 0 \\
\end{array}%
\right)\lambda -
\left(%
\begin{array}{cc}
  \ell & 0 \\
  0 & -E^T\,\ell\,E\\
\end{array}%
\right) +\lambda^{-1}\left(%
\begin{array}{cc}
  0 & \gamma^TF \\
  F^T\gamma & 0 \\
\end{array}%
\right)+\dfrac{c^2}{\lambda^2}\left(%
\begin{array}{cc}
  0 & 0 \\
  0 & F^T\,d\,F \\
\end{array}%
\right)\right]\,,\nn
\en
where
\bq\label{zam}
\gamma=(1+\ell^T)\,k\qquad\mbox{\rm and}\qquad d=k^T\,\ell\,k\,.
\eq
 The matrix
 \[ h=\left(%
\begin{array}{cc}
  0 & 0 \\
  0 & F^T\,F \\
\end{array}%
\right)\equiv\left(%
\begin{array}{cc}
  0 & 0 \\
  0 & r^TF^Tkk^T\,Fr \\
\end{array}%
\right)
\]
is numerical  because $r=1$ (\ref{red-HK}) only.

\begin{prop} The Lax matrix $\hat{L}_y(\lambda)$ (\ref{Lax-hat}) satisfies
(\ref{rpoi}) with numerical $r$-matrix
\bq\label{r-h}
\widehat{r}_{12}^{\,Kow}(\lambda,\mu)=\Bigl[1\otimes\left(1+\dfrac{c^2}{\mu^2}\,h\right)^{-1}\Bigr]
\,\widetilde{r}_{12}^{Kow}(\lambda,\mu)\,
\Bigl[\left(1+\dfrac{c^2}{\lambda^2}\,h\right)\otimes 1\Bigr]\,.
\eq
\end{prop}
The proof is straightforward. Thus, for the generalized Kowalevski
top we constructed third Lax matrix with the numerical $r$-matrix. In
contrast with the previous two Lax matrices in this case we do not
know the physical meaning of the corresponding coordinate system.

The non-zero Poisson brackets between variables $\ell,\gamma$ and $d$
(\ref{zam}) are
\begin{eqnarray}
\{\ell_{ij},\ell_{mn}\,\}&=&
\delta_{in}\,\ell_{jm}+\delta_{jm}\,\ell_{in}-\delta_{im}\,\ell_{jn}-\delta_{jn}\,\ell_{im}\,,\nn\\
\{\ell_{ij},\gamma_{nm}\,\}&=&\delta_{jm}\,\gamma_{ni}-\delta_{im}\,\gamma_{nj}\,,
\nn\\
\nn\\
\{d_{ij},d_{kn}\,\}&=&\delta_{ik}\,d_{jn}+\delta_{jn}\,d_{ik}-
\delta_{in}\,d_{jk}-\delta_{jk}\,d_{in}\label{Poi-Br2}\\
\{d_{ij},\gamma_{nk}\,\}&=&\delta_{in}\,\gamma_{kj}-\delta_{jn}\,\gamma_{ki}
\,,\nn\\
\nn\\
 \{\gamma_{ij},\gamma_{kn}\}&=&c^2\left(\delta_{jn}\ell_{is}-\delta_{ik}d_{jn}\right)\,.\nn
\end{eqnarray}
Entries of matrix $\ell$  jointly with entries of any column of
$\gamma$ form subalgebras with respect to bracket (\ref{Poi-Br2}). If
$\mbox{\rm rank}\, F=1$, then we can put
\[
F^TdF=0, \qquad (\gamma^T\,F)_{ij}=\delta_{j1}\gamma_{i1}
\]
and, therefore, the Lax matrix (\ref{Lax-hat}) is defined on the one
of such subalgebras only.

Let $p=3$, in canonical variables
\[
J_i=-\varepsilon_{ijk}\ell_{jk}\,,\qquad y_j=\gamma_{j1}
\]
the brackets (\ref{Poi-Br2}) coincide  with the standard Lie-Poisson
brackets on  $so(4)$
\begin{equation}\label{bundle}
\,\qquad \bigl\{J_i\,,J_j\,\bigr\}=\varepsilon_{ijk}J_k\,, \qquad
\bigl\{J_i\,,y_j\,\bigr\}=\varepsilon_{ijk}y_k \,,\qquad
\bigl\{y_i\,,y_j\,\bigr\}=c^2\varepsilon_{ijk}J_k.
\end{equation}
Thus, subalgebra $\{\ell,\gamma_{j1}\}$ is isomorphic to the Lie
algebra $so(4)$ \cite{kst03}.

If $q=1$ or $q=3$ the perturbed Hamiltonian (\ref{def-Hg2})
\[H_2=J_1^2+J_2^2+J_3^2-y_1
\]
describes the Lagrange top on the algebra $so(4)$.

For $q=2$ the perturbed Hamiltonian (\ref{def-Hg2})
\[{H}_2=\frac12\Bigl(\,J_1^2+J_2^2+2J_3^2\,\Bigr)
-y_1.
\]
describes the Kowalevski top on the Lie algebra $so(4)$. As for the
usual Kowalevski top \cite{rs03}, we can generalize constraints
(\ref{red-HK}) and describe the Kowalevski gyrostat on $so(4)$
\cite{kst03}.

The Lax matrices for the Lagrange and Kowalevski tops on $so(4)$ were
constructed in  \cite{kst03}. In this section we calculate the
corresponding $r$-matrix (\ref{r-h}) and propose generalization on
the case $p\neq 3$.


\section{Summary}
\setcounter{equation}{0}

We construct the Lax matrices for an integrable deformation of the
known integrable system of two interacting $p$-dimensional and
$q$-dimensional spherical tops. One gets $r$-matrices associated with
this model and with some reduced systems such as generalized Lagrange
and Kowalevski tops.

The matrix $y$ (\ref{y1}), which determines perturbation (\ref{LL})
of the known Lax matrix, is the solution of the equations
(\ref{cond}). Relation of such equations on  the associative algebras
with a general theory of the classical $R$-matrices on the twisted
loop algebras remains an open question. Some another constructions of
such deformations may be found in \cite{golsok}.

The author is grateful to M.A. Semenov-{T}ian-{S}hansky , V.V.
Sokolov and I.V. Komarov for useful discussions. The research was
partially supported by RFBR grants 02-01-00888 .

\end{document}